\documentclass{rnaastex}

\definecolor{black}{rgb}{0,0,0}
\definecolor{red}{rgb}{1.0,0,0}

\newcommand{\UCB}{Department of Astronomy,  University of California: Berkeley, Berkeley, CA 94720}

\newcommand{\NIJ}{Department of Astrophysics/IMAPP, Radboud University, Nijmegen, Netherlands}
\newcommand{\SWIN}{Centre for Astrophysics \& Supercomputing, Swinburne University of Technology, Hawthorn, VIC 3122, Australia}

\newcommand{\SETI}{SETI Institute, Mountain View, California}

\newcommand{\HOU}{Hellenic Open University, School of Science and Technology, Patra, Greece}
\newcommand{\KZA}{University of Malta, Institute of Space Sciences and Astronomy}
\newcommand{\BPF}{The Breakthrough Initiatives, NASA Research Park, Bld. 18, Moffett Field, CA, 94035, USA}
\newcommand{\MIT}{Massachusetts Institute of Technology, Cambridge, MA 02139}
\newcommand{\PENN}{Department of Astronomy and Astrophysics, Pennsylvania State University, University Park PA 16802}

\begin{document}

\title{Breakthrough Listen Follow-up of the Random Transiter (EPIC 249706694/HD 139139) with the Green Bank Telescope}


\correspondingauthor{Bryan Brzycki}
\email{bbrzycki@berkeley.edu}


\author[0000-0002-7461-107X]{Bryan Brzycki}
\affiliation{\UCB}

\author[0000-0003-2828-7720]{Andrew Siemion}
\affiliation{\UCB}
\affiliation{\SETI}
\affiliation{\KZA}

\author[0000-0003-4823-129X]{Steve Croft}
\affiliation{\UCB}

\author[0000-0002-8071-6011]{Daniel Czech}
\affiliation{\UCB}

\author[0000-0003-3197-2294]{David DeBoer}
\affiliation{\UCB}

\author{Julia DeMarines}
\affiliation{\UCB}

\author{Jamie Drew}
\affiliation{\BPF}

\author[0000-0003-2516-3546]{J. Emilio Enriquez}
\affiliation{\UCB}
\affiliation{\NIJ}

\author[0000-0002-8604-106X]{Vishal Gajjar}
\affiliation{\UCB}

\author{Nectaria Gizani}
\affiliation{\UCB}
\affiliation{\HOU}

\author[0000-0002-0531-1073]{Howard Isaacson}
\affiliation{\UCB}

\author{Brian Lacki}
\affiliation{Breakthrough Listen, \UCB}

\author{Matt Lebofsky}
\affiliation{\UCB}

\author{David H.\ E.\ MacMahon}
\affiliation{\UCB}

\author{Imke de Pater}
\affiliation{\UCB}

\author[0000-0003-2783-1608]{Danny C.\ Price}
\affiliation{\UCB}
\affiliation{\SWIN}

\author{Sofia Sheikh}
\affiliation{\PENN}

\author{Claire Webb}
\affiliation{\UCB}
\affiliation{\MIT}

\author{S. Pete Worden}
\affiliation{\BPF}




\section{Introduction}
\label{sec:intro}

The star EPIC 249706694 (HD 139139) was found to exhibit 28 transit-like events over an 87 day period during the Kepler mission's \textit{K2} Campaign 15 \citep{Rappaport:2019}. These events did not fall into an identifiable pattern in arrival times or transit depth, and could not be explained by a multitude of transit scenarios explored by the authors. In addition, there is a nearby star B that is suspected to be gravitationally bound to EPIC 249706694, but it is unclear which star actually hosts these transits. The mystery behind the origin of these events makes this system an interesting target for technosignature searches.

In this note, we analyze observations of EPIC 249706694 taken with the Green Bank Telescope as part of the ongoing Breakthrough Listen search for technosignatures \citep{Worden:2017, Isaacson:2017}. We searched C-band frequencies for Doppler-accelerated narrowband signals. We detect no evidence of technosignatures from EPIC 249706694 and derive an upper limit for the EIRP (Equivalent Isotropic Radiated Power) of $1.0 \times 10^{13}$ W for potential technosignatures.

\section{Observations}
\label{sec:obs}

We observed EPIC 249706694 with the Green Bank Telescope for three 5-minute pointings, starting on 2019 July 03 at 04:30:35 UTC, using the C-band receiver (3564--8439 MHz). We also took 5-minute reference observations between each of these target observations at a $1^\circ$ offset from EPIC 249706694 to reject radio frequency interference (RFI) \citep{Enriquez:2018}. Observations were recorded using the Breakthrough Listen back-end \citep{MacMahon:2018}. In total, 3 on-target observations and 3 off-target observations were used for our signal search analysis. 

For the narrowband signal search, we reduced raw voltage data into high spectral resolution data products with frequency and time resolutions of about 2.79 Hz and 18.25 s, respectively. This is one of the standard archival data products produced as part of the Breakthrough Listen data reduction pipeline \citep{Lebofsky:2019}. All observational data used in this note are available online\footnote{\url{http://blpd0.ssl.berkeley.edu/HD139139/}. The first observation, numbered 0029, is an off-target pointing, after which off- and on-target pointings alternate. The high spectral resolution, high time resolution, and medium resolution data products have suffixes of 0000, 0001, and 0002, respectively.}, including the 3 standard BL data products for each pointing.

Observational and signal search parameters are summarized in Table \ref{params}. We obtained a mean system temperature $T_\textrm{sys}$ of 20.04 K from AutoPeakFocus\footnote{\url{https://science.nrao.edu/facilities/gbt/observing/GBTog.pdf}}. 

\begin{table}
\caption{Parameters for EPIC 249706694 observations and signal search} 
\begin{center} 
\begin{tabular}{ c c } 
    \hline 
    Parameter & EPIC 249706694 \\ \hline \hline
    RA (J2000) & 15$^h$ 37$^m$ 06.2$^s$ \\ 
    Dec (J2000) & -19$^\circ$ 08' 33.1" \\ 
    C-Band Frequency Coverage & 3564--8439 MHz \\ 
    Frequency Resolution & 2.79 Hz \\
    Time Resolution & 18.25 s \\
    Integration Time & 300 s \\
    Initial MJD & 58667.18791 \\ 
    Measured $T_\textrm{sys}$ & 20.04 K \\
    \hline
    Drift Rate Search Limits & $\pm$5 Hz/s \\
    Signal-to-Noise Threshold & 10 \\
    10$\sigma$ Detectable Flux Density Limit & 7.4 Jy \\
    \textbf{Derived 10$\bm{\sigma}$ EIRP limit} & $\bm{1.0\times10^{13}}$ \textbf{W} \\ 
    \hline
    
  \end{tabular} 
  \label{params}
  \end{center}
\end{table}

\section{Results and Discussion}

We used the TurboSETI code \citep{Enriquez:2017} to search for narrowband signals with Doppler drift rates between $\pm 5$ Hz/s and with a signal-to-noise (S/N) threshold of 10. Candidate detections would appear in each of the 3 on-target observations but not in any of the 3 off-target observations. We find that all signals above the S/N=10 threshold fail these criteria; signals appearing in on-target observations also appeared in off-target observations, suggesting that these are anthropogenic in nature. We therefore do not find evidence of technosignatures having S/N$\ge$10 over a 5-minute observing period. 

Based on EPIC 249706694's distance of $107.6\pm0.5$ pc \citep{Lindegren:2018}, our measured $T_\textrm{sys}$ of 20.04 K, and assuming otherwise ideal performance\footnote{\url{http://www.gb.nrao.edu/~fghigo/gbtdoc/sens.html}} at C-band frequencies, we derive a corresponding upper limit for the EIRP of putative transmissions to be $1.0 \times 10^{13}$ W. As a comparison, the Arecibo Observatory's 2380 MHz radar transmitter has an EIRP of about $2\times 10^{13}$ W. We also calculate the minimum detectable flux density to be 7.4 Jy for a transmission with bandwidth 1 Hz, using Eq. 4 in \cite{Enriquez:2017}.

\acknowledgments

Breakthrough Listen is managed by the Breakthrough Initiatives, sponsored by the \href{http://breakthroughinitiatives.org}{Breakthrough Prize Foundation}.  The Green Bank Observatory is a facility of the National Science Foundation, operated under cooperative agreement by Associated Universities, Inc.

\bibliographystyle{aasjournal}
\bibliography{references}

\end{document}